\documentclass[sigconf]{acmart}
\acmSubmissionID{462}

\usepackage{diagbox}
\usepackage{subfig}
\usepackage{multirow}
\usepackage{enumitem}
\usepackage{soul}
\usepackage[skip=3pt]{caption}

\setcitestyle{square} %

\newcommand{\oslot}{\texttt{out}}
\newcommand{\islot}{\texttt{in}}

\copyrightyear{2023}
\acmYear{2023}
\setcopyright{rightsretained}
\acmConference[SIGGRAPH '23 Conference Proceedings]{Special Interest Group on Computer Graphics and Interactive Techniques Conference Conference Proceedings}{August 6--10, 2023}{Los Angeles, CA, USA}
\acmBooktitle{Special Interest Group on Computer Graphics and Interactive Techniques Conference Conference Proceedings (SIGGRAPH '23 Conference Proceedings), August 6--10, 2023, Los Angeles, CA, USA}
\acmDOI{10.1145/3588432.3591520}
\acmISBN{979-8-4007-0159-7/23/08}

\begin{document}

\author{Yiwei Hu}
\affiliation{
	\institution{Yale University}
	\city{New Haven}
	\state{CT}
	\country{USA}
}
\affiliation{
	\institution{Adobe Research}
	\city{San Jose}
	\state{CA}
	\country{USA}
}
\email{yiwei.hu@yale.edu}

\author{Paul Guerrero}
\affiliation{
	\institution{Adobe Research}
	\city{London}
	\country{UK}
}
\email{guerrero@adobe.com}

\author{Miloš Hašan}
\affiliation{
	\institution{Adobe Research}
	\city{San Jose}
	\state{CA}
	\country{USA}
}
\email{mihasan@adobe.com}

\author{Holly Rushmeier}
\affiliation{
	\institution{Yale University}
	\city{New Haven}
	\state{CT}
	\country{USA}
}
\email{holly.rushmeier@yale.edu}

\author{Valentin Deschaintre}
\affiliation{
	\institution{Adobe Research}
	\city{London}
	\country{UK}
}
\email{deschain@adobe.com}
\renewcommand{\shortauthors}{Hu et al.}

\title{Generating Procedural Materials from Text or Image Prompts}

\begin{teaserfigure}
\centering
\includegraphics[width=\textwidth]{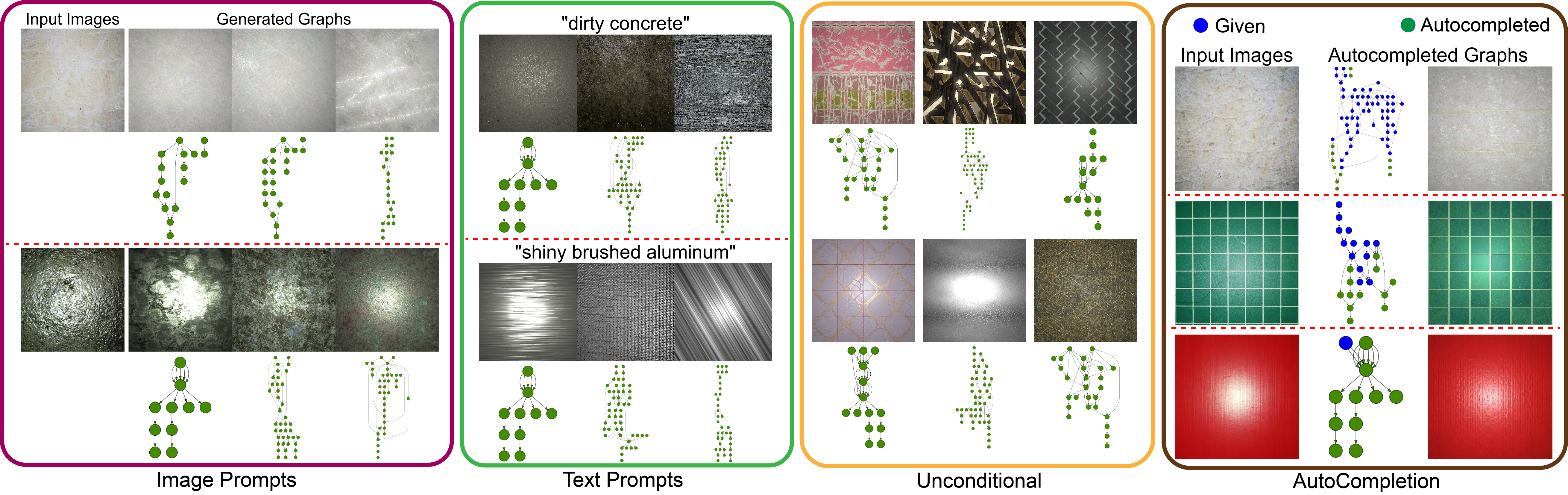}
\caption{Procedural materials can be represented by directed computational graphs where each node represents a 2D image generator or filtering operator. The nodes are connected by unidirectional edges defining the computation flow. Our generative model can produce multiple procedural material graphs 1) from image prompts, 2) from text prompts, 3) unconditionally and 4) conditioned on partial graphs (AutoCompletion). 
Generated graph structures are shown in green, existing structures (in AutoCompletion) are in blue.}
\label{fig:teaser}
\end{teaserfigure}
\begin{abstract}
Node graph systems are used ubiquitously for material design in computer graphics. They allow the use of visual programming to achieve desired effects without writing code. As high-level design tools they provide convenience and flexibility, but mastering the creation of node graphs usually requires professional training. 
We propose an algorithm capable of generating multiple node graphs from different types of prompts, significantly lowering the bar for users to explore a specific design space. Previous work~\cite{Guererro22} was limited to unconditional generation of random node graphs, making the generation of an envisioned material challenging. We propose a multi-modal node graph generation neural architecture for high-quality procedural material synthesis which can be conditioned on different inputs (text or image prompts), using a CLIP-based encoder. We also create a substantially augmented material graph dataset, key to improving the generation quality. Finally, we generate high-quality graph samples using a regularized sampling process and improve the matching quality by differentiable optimization for top-ranked samples. We compare our methods to CLIP-based database search baselines (which are themselves novel) and achieve superior or similar performance without requiring massive data storage. We further show that our model can produce a set of material graphs unconditionally, conditioned on images, text prompts or partial graphs, serving as a tool for automatic visual programming completion. 
\end{abstract}




\maketitle

\section{Introduction}

Node graph systems are widely adopted in the computer graphics industry, allowing artists to design various assets such as material or shader graphs.
This kind of system provides a more user-friendly visual design interface than shading languages, while still offering expressive power. In this work we focus on the \emph{material node graphs} which are commonly used in material authoring~\cite{SubstanceDes}. These graphs describe a set of material maps using a combination of noise and pattern generators and filters. Procedural materials have attractive properties like tileability, arbitrary resolution and convenient parametric editability. However, high-quality node graphs are difficult to author, requiring significant expertise and time from artists. 

In recent years, procedural material fitting has seen significant progress, with parameter estimation and optimization methods~\cite{hu2019, Shi20, hu2022diff} and a segmentation-based graph fitting framework~\cite{hu2022inverse}. These methods however rely on either a small fixed set of available graphs or a generic graph structure instantiated by user segmentation. Most recently, MatFormer~\cite{Guererro22} was proposed, enabling unconditional graph generation through multiple transformers~\cite{vaswani2017transformers}. While useful for random material exploration, it can not be guided by a specific target appearance. In this work, we propose a novel multi-modal conditional generative model 
for material graphs. The model is  multi-modal in that it can work with no user input, a user provided image, a user provided text prompt, or a partial node graph as input. Our transformer model can 
produce a variety of graphs and enable automatic graph completion (Fig. \ref{fig:teaser}).

In particular, we present a CLIP-based \cite{CLIP} encoder module to enable a multi-modal conditioning on either text or image prompts. Each of the transformer layers is conditioned by a CLIP embedding mapped by learnable MLPs. 
To train our conditional generative model we curate a new material graph dataset from Substance Source \cite{SubstanceDes}. Given the limited 
available data, 
we augment the data and improve numerous details in the graph representation. 
This is particularly important not only for conditional generation to reproduce a target appearance, but also improves unconditional generation.

Further, we propose validation and regularization steps at inference time to ensure high-quality error-free graph generation. To account for the difference between the visual error and the parameter space error --on which our transformers are trained-- and improve the image-space matching quality, we further improve the top-ranked generated graphs through differentiable optimization~\cite{Shi20}.

To evaluate the performance of our conditional model, we present two CLIP-search-based baselines. We show our conditional generative model achieves similar or better performance compared to retrieval from a large pre-generated database, but without the massive storage footprint. 
We  present applications of our method for text-conditioned generation and conditional automatic graph completion, modes not supported by previous work in modeling.

In summary we present:
\begin{itemize}
    \item A multi-modal conditional architecture for material graph generation.
    \item A %
    graph dataset augmentation and cleanup strategy.
    \item A sampling regularization and post-sampling procedure to minimize the image/text-space distance with the generated graphs.
\end{itemize}

\section{Related Work}
\subsection{Program and Graph Generation in Graphics}
Node graph systems are visual programming interfaces, making the generation of a node graph akin to program 
synthesis. %
 In computer graphics, Hu et al. \shortcite{hu2018exposure} and Ganin et al. \shortcite{ganin2018synthesizing} synthesized programs for interpretable image editing by predicting a sequence of image processing operations using reinforcement learning. Other work in program synthesis focused on 2D or 3D procedural shape generation. For example, Stava et al. \shortcite{vst2010inverse} proposed a framework to automatically generate L-systems while Demir et al. \shortcite{demir2016proceduralization} extracted a context-free parameterized split grammar for 3D shapes. Following the development of learning-based approaches, recent research \cite{du2018inversecsg, NEURIPS2019_50d2d226, NEURIPS2018_67880768, Sharma_2018_CVPR, tian2018learning, lu2019neurally, jones2020shapeAssembly, johnson2017inferring, walke2020learning, kania2020ucsgnet, xu2021inferring, wu2019carpentry} modeled priors over shape programs which can either be used for program generation or program induction from input shapes. %

We build on the recent work of Guerrero et al. \shortcite{Guererro22} which presented a transformer-based unconditional generative model %
for material graphs. We improve its unconditional sampling %
to enable conditional generation. For multi-modal conditioning, we rely on the joint image/text embedding of CLIP \cite{CLIP} which has been shown to be effective in
text-to-image~\cite{StableDiffusion, Dalle2} and texture synthesis~\cite{Song22} applications.

\subsection{Inverse Procedural Material Modeling}
Our method 
is an inverse procedural material modeling approach when conditioned on an image. Procedural materials define materials or texture maps as a set of procedures \cite{Liu16, Guehl20}, which are typically organized as a node graph for interaction flexibility \cite{SubstanceDes}. Given these existing procedures, different inverse methods \cite{GuoBayesian20, hu2019, Shi20, hu2022diff} aim to recover parameters for a predefined procedural material model given a target image.
More recently, Hu et al. \shortcite{hu2022inverse} go beyond parameter regression, attempting to synthesize the structure of a node graph instead of relying on a fixed node graph fetched from a database. The method however still relies on a generic graph structure, specified by a user segmentation of the material. 
Our conditional generative model, on the other hand, %
can generate various node graph structures given different types of user input i.e., text, images or partial graphs. %

Further, previous work in inverse modeling has focused on the production of a single node graph reproducing the details of a given material exemplar. Our model can generate multiple node graphs, following the prompt in terms of semantics, structures, and colors, rather than precisely matching details. The user can then select from the set of produced node graphs to continue to explore the material design space.

As we synthesize computational graphs, our method also differs from image-based inversion e.g., StyleGAN inversion \cite{richardson2021, Tov2021, Guo20}. Indeed, our results benefit from the advantages of procedural representation for controlability, tileability and arbitrary resolution.

\subsection{Material Acquisition and Generation}
We approach a problem similar to material acquisition that recovers material maps from images.
However, material acquisition attempts to accurately estimate material properties in the form of texel values of material maps from one or more measurements (generally photographs). Classical material reconstruction \cite{Guarnera16} requires dense measurements, with tens to thousands of captured images to accurately digitize the material optical properties of a given target. Recent methods based on deep learning use a large amount of synthetic material training data to present various lightweight material acquisition frameworks, reducing the number of photos required to less than ten, with many requiring a single input flash photograph \cite{Li17, Deschaintre18, Deschaintre19, Zhou21, Guo21, Ye21, Gao19, Guo20, henzler21neuralmaterial, zhou22}.%

As opposed to these methods, our goal is not to directly produce an array of texels representing the material.
Rather, we generate programmable procedural node graphs. The generated node graphs can then generate material parameter maps at any resolution, with editable sub-components for convenient manipulation and variation.%

\section{Conditional Material Graph Generation}

\subsection{Overview}

We present a generative model for procedural materials that can be conditioned on a given text prompt or image. Procedural materials are represented as node graphs, which are directed acyclic computation graphs (Fig. \ref{fig:teaser}) that can be controlled through a set of parameters, and output a set of 2D material maps that define a material. A node graph consists of nodes that represent image operators, and edges that define the information flow between operators.

Our goal is to model the conditional distribution $p(g|y)$ of graphs $g$ given inputs $y$, which may, for example, be text or image prompts. To model the probability distribution over node graphs, we follow MatFormer~\cite{Guererro22} and encode 
the graph as a set of sequences: a \emph{node sequence} $S_n$, an \emph{edge sequence} $S_e$ and a \emph{parameter sequence} $S_p$ that are obtained by linearizing the graph. A probability distribution over each sequence can then be modeled using three transformer-based models~\cite{vaswani2017transformers}, for nodes, edges, and parameters. We model correlations between the three sequences by conditioning edge generation on nodes and parameter generation on both nodes and edges.
Unlike MatFormer, we also condition the generation of all three sequences on the inputs $y$, and make several architectural improvements to the generators.

We describe node graphs in Sec.~\ref{Sec:node_graphs} and give a more detailed description of our conditional generative models in Sec.~\ref{Sec:cond_gen}. To train our conditional model, we present a new material graph dataset 
in Sec.~\ref{Sec:dataset}, with careful preprocessing and extensive data augmentation, significantly extending the material space captured by our model compared to MatFormer. We describe our training setup in Sec.~\ref{Sec:training} and discuss our regularized sampling as well as  ranking-and-optimize step for image-space similarity improvement in Sec.~\ref{Sec:sampling}.

\subsection{Node Graphs}
\label{Sec:node_graphs}
A material node graph $g = (N, E)$ consists of \emph{nodes} $N=(n_1, n_2 \dots)$ and \emph{edges} $E=(e_1, e_2, \dots)$.
Nodes are instances of image operations from a predefined library. A node $n_i = (\tau_i, P_i)$ is defined by the operation type $\tau_i$ and a set of node parameters $P_i = (p^i_1, p^i_2, \dots)$ that control the operation. Individual node parameters $p^i_j$ may have various types like integers, floats, strings, fixed-length arrays or variable-length arrays. The operation type also defines a set of input slots $(\islot^i_1, \islot^i_2, \dots)$ and a set of output slots $(\oslot^i_1, \oslot^i_2, \dots)$. Each input slot can receive one input image, and the set of all input images is transformed by the image operation into output images that are provided in the output slots. A node that has zero input slots is called a \textit{generator node}. All nodes have at least one output slot, except special \emph{output nodes} that define the final outputs of the graph.

Edges %
define the information flow in a node graph. They are unidirectional: an edge $e_i = (\oslot^j_l, \islot^k_h)$ always connects an output slot of one node to an input slot of a different node. An output slot can provide images to multiple input slots of other nodes, but an input slot can only receive an image from a single output slot. Additionally, no cycles are allowed in a node graph. In the following, we denote the output slot that edge $e_i$ starts from as $e_i^{\oslot}$ and the input slot that the edge ends in as $e_i^{\islot}$.

A graph is evaluated by running the operations defined by each node in a topological order. After evaluation, the output nodes of the graph provide a set of 2D material maps describing spatially varying material parameters such as diffuse color, roughness, height, or normals.

\begin{figure}
\centering 
\includegraphics[width=1.0\linewidth]{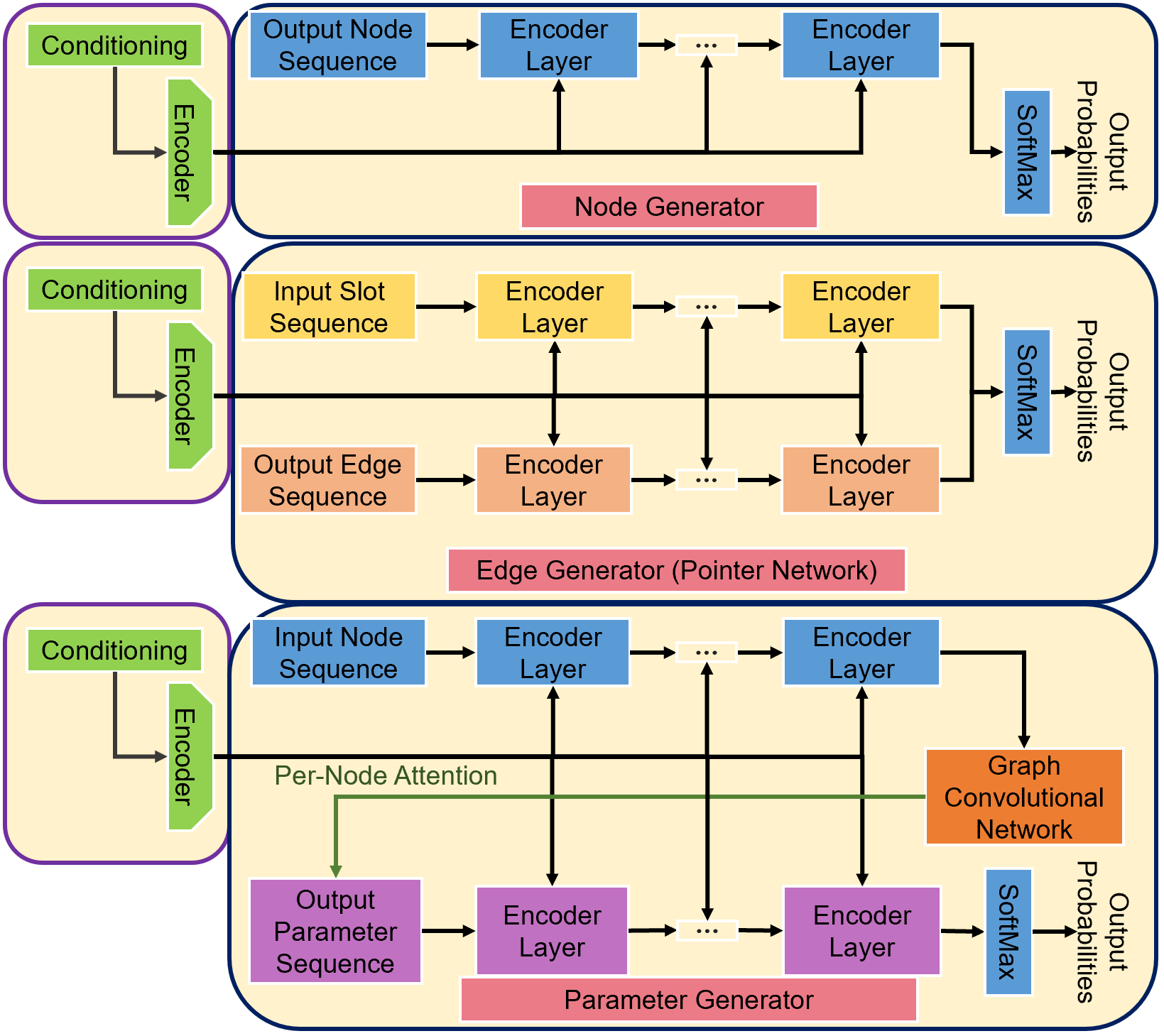}
\caption{The architecture of our conditional generative model. We show a single next token generation step. The "Output Node/Edge/Parameter Sequence" blocks are the sequences generated so far. As an autoregressive model, the output sequences serve as inputs for next token prediction. Given a text or image prompt as conditioning, the encoder encodes it to the dimension of the hidden states of our transformer encoders. For the encoder, we use CLIP embedding with a learnable MLP. The encoded feature vector is then fused into each transformer encoder layer. In the parameter generator, the Graph Convolutional Network captures the edge connectivity of neighbor nodes and the transformed node embedding is attached as an auxiliary input sequence for the transformer to pay attention to the node that it is generating parameters for.}
\label{fig:network}
\end{figure}
\subsection{Conditional Generative Model} 
\label{Sec:cond_gen}
Our generative model for node graphs is based on MatFormer,
which we improve to allow for generation \emph{conditioned} on images or text prompts.
We generate node graphs in three steps, corresponding to nodes, edges, and node parameters. For each step, we train a conditional transformer that models the probability distribution over node, edge, or parameter sequences, respectively, conditioned on the input prompt $y$. A transformer-based conditional generative model
with parameters $\xi$ models the conditional probability distribution
of a sequence $S$ conditioned on $y$ as a product of conditional probabilities for each token $s_i$:
\begin{equation}
    p_\xi(S|y) \coloneqq \prod_i p_\xi(s_i|s_{<i}, y),
\end{equation}
where $s_{<i} := s_1, ..., s_{i-1}$ denotes a partial token sequence generated up to token $s_i$. The transformer outputs the probability distribution $p_\xi(s_i|s_{<i}, y)$ in each step, which can be sampled to obtain the next token $s_i$. The dependence between the sequences that represent a graph $g$ is modeled using conditional probabilities:
\begin{equation}
p(g|y) \coloneqq p_\theta(S_n | y)\ p_\phi(S_e | S_n, y)\ p_\psi(S_p | S_e, S_n, y),
\end{equation}
where $\theta$, $\phi$, and $\psi$ are parameters of the transformer models for node sequences $S_n$, edge sequences $S_e$, and parameter sequences $S_p$, respectively. 
Unlike MatFormer, our generative model accepts a condition $y$ that guides the sampling process of the transformer. The condition $y$ is given as a feature vector. We will describe our approach to obtain this feature vector from images or text prompts, and our transformer conditioning strategy in Sec.~\ref{Sec:cond_encoding}.  In a further departure from MatFormer, we generate all parameters in a graph as a single sequence, instead of one sequence per node, and condition the parameter probabilities $p_\psi(S^i_p | S_e, S_n, y)$ on the edge sequence $S_e$ in addition to the node sequence $S_n$ (Sec. \ref{sec:param_gen}). %

\subsubsection{Conditioning on Text Prompts or Images}
\label{Sec:cond_encoding}
We consider multi-modal inputs including text prompts and images. Radford et al. (CLIP)~\shortcite{CLIP} proposed a jointly learned space for encoding both text prompts and images. To benefit from this joint encoding, we use a frozen CLIP model (ViT-L/14) to encode our inputs. As we will introduce in Sec. \ref{Sec:dataset}, we train our conditional models using image-to-graph correspondences. During training, our network is only conditioned on image space CLIP embeddings which are not exactly the same as CLIP text space embeddings \cite{Dalle2}. This limits the quality of the text-conditioned generated material graphs. We therefore follow the approach of DALL·E2 and apply a prior (a specialized network) to transform CLIP text embeddings into CLIP texture embeddings. This model was trained using 10 millions of text/texture pairs~\cite{aggarwal2023controlled}, transforming text embedding into texture image embeddings.
When providing a text prompt, we encode it as a CLIP text embedding and then transform it to a CLIP texture image embedding, compatible with our training data. The CLIP embedding is then encoded by a trainable MLP to map it to the dimension of the hidden states of our transformer encoders. The mapped embedding is then added to the output of each transformer encoder layer, after layer normalization. We show in Fig. \ref{fig:network} the design of our conditional generative model. %

Conditioning on CLIP embedding allows our model to accept both image and text prompts as inputs. Other encoding methods are possible. We experimented with an alternative image encoding approach. While CLIP captures the high-level semantic information in an image, to encode low-level texture statistics 
we add VGG features statistics to capture fine-scale texture detail and a 16x16 downsampled thumbnail to summarize the main color of the input image. Detailed implementation of this encoding is presented in the supplementary material. We discuss the performance of this alternative (\textbf{Ours~(VGG)}) in Sec. \ref{sec:results}. Note that the construction of this encoding is however slightly more complicated and prevents the use of text encoding. With the exception of the comparison in Fig.~\ref{fig:comp}, the image results we show in the paper are generated by our CLIP-only encoding (\textbf{Ours}).

\subsubsection{Node and Edge Generation}
Node and edge generation closely follows the approach proposed by MatFormer, %
except for the added conditioning on $y$. 

To generate a \emph{node sequence} $S_n = (\tau_0, \tau_1, \dots)$, we  iteratively sample the node model $p_\theta(s^n_i |s^n_{<i}, y)$, where $s^n_i$ denotes element $i$ of sequence $S_n$. Each step generates the integer ID $\tau_i$ of a node.

To generate an \emph{edge sequence} $S_e = (e_1^\oslot, e_1^\islot, e_2^\oslot, e_2^\islot, \dots)$, we iteratively sample the edge model $p_\phi(s^e_i |s^e_{<i}, S_n)$. Each step generates a pointer into the list of all output and input slots in the graph. Pointers are generated using a transformer with a head based on Pointer Networks~\cite{vinyals2015pointer}. The list of all output and input slots is derived from the node sequence $S_n$ generated in the previous step and includes information about the operation type of the node each slot is attached to.

As usual for transformers, all sequences are extended 
with auxiliary tokens that mark the start and end of a sequence. Additionally, all transformer models receive auxiliary input sequences that provide information such as the index of a token in the sequence. We use the same auxiliary sequences as MatFormer,
please refer to the supplementary material for details. %

\subsubsection{Parameter Generation}
\label{sec:param_gen}
For parameter generation, we take a different approach than MatFormer. First, we condition on both the node and edge sequences $S_n$ and $S_e$ instead of only on the node sequence $S_n$. This allows us to capture the relationship between parameters and edges in the graph; these were independent in MatFormer.
Second, instead of generating one parameter sequence per node, we generate all parameters of a graph in a single sequence. %

To generate the \emph{parameter sequence} $S_p = (p^1_1, p^1_2, \dots, p^2_1, p^2_2, \dots)$ of a graph, we iteratively sample 
the 
model $p_\psi(s^p_j | s^p_{<j}, S_n, S_e)$. Each step outputs a parameter, or one scalar element of a parameter in the case of vector- or array-valued parameters. Conditioning on the node and edge sequences $S_n$ and $S_e$ is implemented through a context-aware embedding $\overline{\tau}_i$ of each node $n_i$. These node embeddings are given as an auxiliary input sequence, where each parameter token receives the embedding of the node it is being generated for (i.e., per-node attention in Fig. \ref{fig:network}).

The embedding $\overline{\tau}_i$ includes information about both the operation type $\tau_i$ of the node and the full node sequence $S_n$. It is computed with a transformer encoder $\overline{\tau}'_i = g^p(S_n, i)$. We additionally include edge connectivity information from the sequence $S_e$ in the node embedding. We use a graph convolution network (GCN) to capture the edge connectivity in the neighborhood of the node: given a node embedding $\overline{\tau}'_i$, we use a GCN $h$ with a residual connection to capture the local edge connectivity information:
\begin{equation}
    \overline{\tau}_i = \overline{\tau}'_i + h(S_n, S_e, i).
\end{equation}
We use 6 layers in our GCN, allowing us to capture the edge structure up to 6 edges away from the node $n_i$. %

The parameter sequence is extended with auxiliary start and end tokens, and with special tokens that mark the start of the parameter sub-sequences for each node. Apart from the auxiliary sequence of node embeddings $\overline{\tau}_i$, we use the same auxiliary input sequences as MatFormer, providing information such as the index and type of the parameter being generated. We also compare our graph-conditioned parameter generator with an image/text conditioned extension of the MatFormer parameter generator. We find a small performance improvement and training speed acceleration (1.5x) due to higher level of parallelism. See our supplementary material for a discussion.

\subsection{Material Graph Dataset} 
\label{Sec:dataset}
The success of sequence generative models such as GPT3 \cite{brown2020language} relies largely on the size of their training data. However, the data available for procedural material graphs is limited. To generate a dataset which can be used to train our conditional model, we rely on Substance Source \cite{SubstanceDes}. However, the available material graphs were created by different material artists over multiple years, with different methodology and skills, making the data inconsistent. We therefore carefully preprocess the dataset, and perform a series of data augmentations. 
\subsubsection{Graph Reformatting and Simplification} Since the raw dataset contains material graphs created in different system versions,
our first step is to ensure that all material graphs are properly formatted and have uniform material representation e.g., fixing precision and dependency problems (See supplementary material for a checklist).

Second, a material graph can generate different material parameter maps for different rendering workflows %
and can be designed to generate additional material maps such as ambient occlusion (AO) maps, anisotropic maps etc. Such parameter maps are useful for artists but make the conditional generative task particularly difficult, as not all graphs contain these outputs and they tend to increase the length of the sequence to generate. Hence, we focus on the standard physically-based-rendering (PBR) workflow, assuming a GGX microfacet \cite{Walter07} shading model, and generate node graphs that produce albedo, normal, roughness and metallic maps. We therefore prune the branches of the unused maps from the graph where they appear, in a back-to-front way. A node is discarded when it is in the computation branch of a unused node, and not in the computation of a used node. This simplification step significantly reduces the size of graphs and lets our model focus on the most important sequences. To avoid the generation of biased normals sometimes observed in MatFormer, we also prune the Level nodes (the nodes that adjust the histogram of the input image) in the normal branch.

\subsubsection{Graph Splitting and Filtering}
In material graph design systems, switch nodes are a type of node that activates one of the multiple alternative computation branches passing this node. The branch being activated is controlled by a integer parameter specified by users. This essentially packs multiple functionalities into a single graph and %
is extremely difficult for our model to learn. A graph that contains switch nodes can be split into multiple smaller graphs by creating a version of the graph for each branch, further reducing its complexity. If multiple switch nodes are present, a large number of combinatorial options may exist. We set a sampling upperbound to $5$. If the maximum number of branches is $k_b$ among all the switch nodes, we sample at least $\max(k_b, 5)$ to ensure each branch is at least sampled once. Finally, we remove graphs which are too similar to others, to prevent duplicates (i.e. the average mean square difference in the material maps they generate is smaller than 0.01).

Considering the difficulty of predicting overly long sequences, we cut the graphs belonging to the tail of the length distribution. We filter node graphs which contain a large number of nodes (> 80) or edges (> 200) or slots (> 210). This graph reduction step ensures our network focuses on statistically well represented graph structures. Our filtered dataset includes $4667$ ($72\%$ of our unfiltered dataset) valid graph instances, compared against the $2820$ unfiltered instances in MatFormer.
This filtering is particularly important to our conditional model which has to match a specific desired appearance.

\subsubsection{Parameter Augmentation}
\label{Sec:param_aug}
The parameters of each node are crucial to control the behavior of a given graph. To increase the number of variations of materials available during the training of our generator, we sample 100 sets of parameters for each of the $4667$ graphs of our dataset. 

However, randomly sampling the full parameter space of a procedural model will generate incorrect parameter combinations, producing purely white or black images or material maps which are not useful to artists~\cite{hu2022diff}. We compute the statistical distribution on parameters from the Substance Source dataset to guide our sampling process. Specifically, we sample a parameter $p$ in a material graph $g$ based on a Gaussian distribution $G(\mu^g_p, \beta \sigma_p)$. The mean of this Gaussian is the default value $\mu^g_p$ defined in the graph $g$'s preset and the standard deviation is scaled by a factor $\beta$ from $p$'s standard derivation among the whole dataset $\sigma_p$. Some parameters' statistics are not reliable due to limited observations in the dataset. In such case, we use an uniform distribution based on a scaled range i.e., $U((1-\alpha)\mu^g_p, (1+\alpha)\mu^g_p)$ by a scaling factor of $\alpha$. The $\alpha$ and $\beta$ are empirically selected to achieve a balance between fidelity (i.e., the sampled graphs still look like a material) and diversity. $\alpha$/$\beta$ are set to 0.06/0.2 when sampling float types and 0.06/0.5 for integer types.

\begin{figure} [t]
\centering
\addtolength{\tabcolsep}{-4pt}
\begin{tabular}{ccc}
\includegraphics[width=0.15\textwidth]{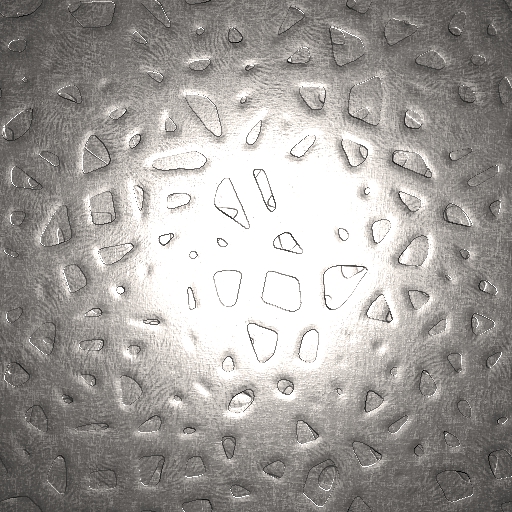} &
\includegraphics[width=0.15\textwidth]{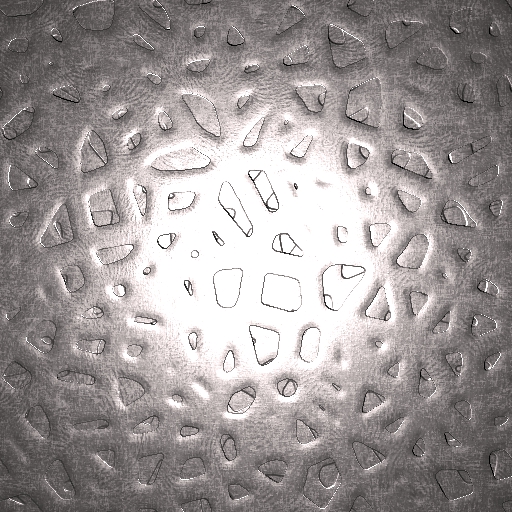} &
\includegraphics[width=0.15\textwidth]{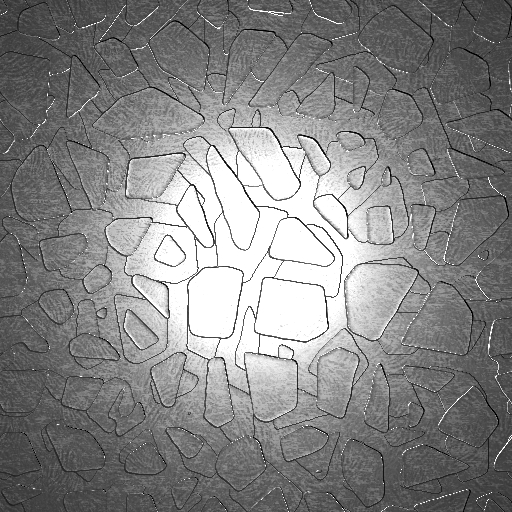} \\
Ground Truth & 128 Quantization & 32 Quantization
\end{tabular}
\caption{Using 32 bins for parameter quantization results in significant reconstruction error. A higher number of quantization bin (128 here) allow to better reconstruct the original appearance.}
\label{fig:quant}
\end{figure}

\subsubsection{Final Dataset}
Our final improved dataset contains 466,700 cleaned graphs that are more amenable to generation, paired with their associated output material maps. %
We render these material maps on a planar surface with a point light collocated with the camera, to synthesize an image-graph pair for training and evaluation. (We could use other lighting configurations like environment mapping, since CLIP encoding is fairly insensitive to precise lighting.) For parameter generation, the probability is modeled over quantized values. We use 128 quantization bins (compared to the 32 used by MatFormer). While more quantization bins are more challenging to learn, the result shows significantly better reconstruction quality as displayed in Fig. \ref{fig:quant}. This is particularly important for our model to match the user-provided condition.
 
\subsection{Training}
\label{Sec:training}
We train our conditional generative model with our new dataset. We split the dataset into a training set and a validation set before parameter augmentation, to ensure that the training set and validation set contain graphs with different topologies, not just different parameter settings. 

All three models (node transformer, edge transfer and parameter transformer) are trained with the ground truth graphs as supervision, using a binary cross-entropy loss over the probabilities estimated by the transformer generators. Each transformer is trained separately using teacher forcing: that is, when generating a new token in a sequence, the ground-truth sequence is used as previously generated tokens. As we have a limited number of different graph structures in the dataset, we ensure we do not over-fit by keeping the checkpoint with the minimum validation loss.

\subsection{Sampling, Ranking and Optimization} 
\label{Sec:sampling}
\subsubsection{Sampling}
Since our generative models consist of three transformers operating in succession, errors from previously generated sequences can propagate and affect the quality of sequences dependent on them. For this reason, we carefully decode the sampled sequences using semantic validity checks to ensure error-free generation. 

While MatFormer ensures semantic validity as a post-processing step, once the entire sequence has been generated,  we perform the validity checks during sampling, making sure to choose only among semantically valid choices for any given token. This makes sure that the chosen tokens are consistent with the semantically valid choices for the previously generated tokens. See our supplementary material for the detailed validation rules we apply.

After node and edge generation, an uninitialized graph is ready for parameter prediction (Fig. \ref{fig:network}). The graph structure serves as a condition to the parameter sequence. Considering possible prediction errors, for robust parameter generation, we regularize the generated graph by removing unconnected nodes.

\subsubsection{Ranking and Optimization}
\label{sec:optim}

During inference, we sample each token according to the probability distribution predicted by the transformer until reaching the end tokens. However, the quality of generated graphs is hard to evaluate. The cross-entropy loss of an estimated probability for a token does not directly reflect the true appearance distance to the input, since distance in the parameter space does not necessarily reflect distance in the image/text space. We would like to measure the discrepancy between the sampled graphs and the input prompts in the same space. We therefore de-serialize the predicted token sequences to reconstruct a material graph. Evaluating this material graph generates material maps we render. We can then use a CLIP cosine metric to measure the difference between our graph's results and the input text/images (alternatively, a sliced Wasserstein distance \cite{Heitz2021, hu2022control} can be used to measure the statistical distance to the input images). 
\begin{table}
\caption{Statistical results expressed as style loss. Unoptimized: searched or our predicted results. Optimized: Differentiable optimized results using style loss. We searched/predicted 30 samples and optimized 10 of them. When computing statistics, we count for the top-5 samples after ranking (by style loss) in order to remove outliers. The style loss we used here as metric is the L1 difference of Gram Matrices of VGG features plus L1 difference of 16x16 downsampled thumbnails (weighed by 0.1).  We report Best-of-5: the minimum loss among the top-5 samples and Average-of-5: the average loss among the top-5 samples. \textbf{Lower is better}. Our model can achieve statistically better or similar performance comparable to search in a giant pre-generated image database. We also include additional quantitative comparisons to a class-conditioned generator which has a loss of 0.0544 (Best-of-5) and 0.0774 (Avg-of-5) before optimization. See our supplementary material for details.}
\centering
\begin{tabular}{|ccccc|}
    \hline
    Unoptimized & Ours & Ours~(VGG) & Ours Uncond & Dataset \\
    \hline
    Best-of-5 & 0.0314 & \textbf{0.0300} & 0.0382 & 0.0384\\
    \hline
    Avg-of-5 & 0.0346 &  \textbf{0.0329} & 0.0539 & 0.0499\\
    \specialrule{2.5pt}{1pt}{1pt}
    Optimized & Ours & Ours~(VGG) & Ours Uncond & Dataset \\
    \hline
    Best-of-5 & 0.0218 & 0.0199 & 0.0194 & \textbf{0.0191}\\
    \hline
    Avg-of-5 & 0.0242 & \textbf{0.0221} & 0.0227 & 0.0237\\
    \hline
\end{tabular}
\label{table:stats}
\end{table}
A differentiable optimization step \cite{Shi20, hu2022diff} can be further applied to refine the accuracy of predicted parameters to better match the given inputs. We generate multiple graph samples and rank them based on the text/image space metrics (CLIP cosine or sliced Wasserstein) and pick the top-k for further optimization (k=5 in our experiments). As node graphs are not all differentiable, we limit the refinement to the optimizable nodes only.
Despite differentiable optimization limitation to adjust parameters only --it cannot modify the generated graph topology-- this step helps reduce the prediction appearance error, especially in terms of color and roughness values as shown in Fig. \ref{fig:image_inputs}. For each image or text prompt, multiple graphs can be synthesized. The inference takes around 1.32 seconds \textit{per graph}. If performing optimization, it takes $2\sim3$ minutes \textit{per graph}, measured on a NVIDIA RTX 3090.

\section{Implementation Details}
Our transformer-based generative models are built upon the GPT \cite{brown2020language} architecture. We select slightly different hidden dimensions, layers, and heads for each transformer to adjust for the size of our dataset (40/2/4 for node generator; 48/2/4 for edge generator; 96/4/4 for parameter generator). Regarding node order, we follow MatFormer to encode the node sequences in a back-to-front breadth-first traversal order ${\pi}_r$. For the autocompletion task, we use reversed ${\pi}_r$ i.e. from last to first node of ${\pi}_r$.

We train our models with the Adam optimizer \cite{Adam}, using a learning rate of $1e^{-5}$ and a batch size of 64. We train each transformer model in parallel on three NVIDIA V100 GPUs. The total training time for each transformer is $\sim$22 hours for node generator; $\sim$28 hours for edge generator and $\sim$36 hours for parameter generator. During the sampling phase, we apply nucleus filtering with a top-p value of 0.9 and sample the top-5 candidates without adjusting the softmax temperature.

\section{Results}
\label{sec:results}
We show our conditional generative model can generate material graphs for text or image prompts. We also show improved variety for unconditional generation and demonstrate automatic conditional graph completion. Over 90\% of generated graphs are valid. Not all generated graphs may fit the desired semantics, but generation is very fast, letting us quickly sample multiple graphs and automatically select the top-ranked ones.

To quantitatively evaluate the performance of our model, we compare it to two novel, challenging, CLIP-searched-based baselines:
\begin{itemize}
    \item Searching and optimizing in a database of 102,400 pre-sampled material graphs generated using the unconditional version of our model, trained on our new augmented dataset. (\textbf{Ours Uncond})
    \item Searching and optimizing directly in our new augmented dataset containing 466,700 graph variations. (\textbf{Dataset})
\end{itemize}
We create a test image dataset containing 48 real photos. Given an input photo, we retrieve the 30 closest samples in the different databases in CLIP space and optimize (using the MATch framework~\cite{Shi20}) the top-5 samples. We additionally optimize the top-5 (out of the 30 retrieved) samples with lowest visual statistical difference (sliced Wasserstein distance~\cite{Heitz2021}), for a total of 10 optimized samples. For our method, we first conditionally generate 
150 graphs and perform the same search-and-optimization approach. Our conditional generative model achieves better or similar performance as shown in Table \ref{table:stats}. We note that our proposed alternative conditional encoding adding VGG features (\textbf{Ours~(VGG)}) produces slightly lower numerical error, but does not allow for text encoding. Visually, our results, our alternative encoding, and baselines are close, as shown in Fig. \ref{fig:comp} (more visual comparisons and detailed statistical analysis with precise definition of style metrics can be found in the supplemental material). However, %
the storage required of our model is orders of magnitude smaller than the pre-generated database: only $15$ MB against $115$ GB. Further, our model enables additional applications such as autocompletion (Fig. \ref{fig:autocompletion}). 

\begin{figure}
\centering
\addtolength{\tabcolsep}{-4pt}
\begin{tabular}{cccc}
\includegraphics[width=0.11\textwidth]{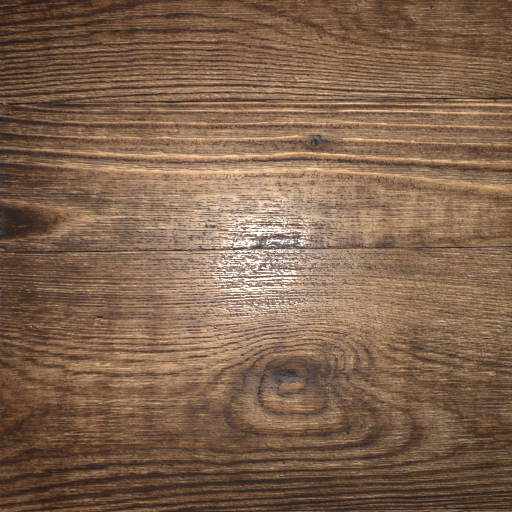} &
\includegraphics[width=0.11\textwidth]{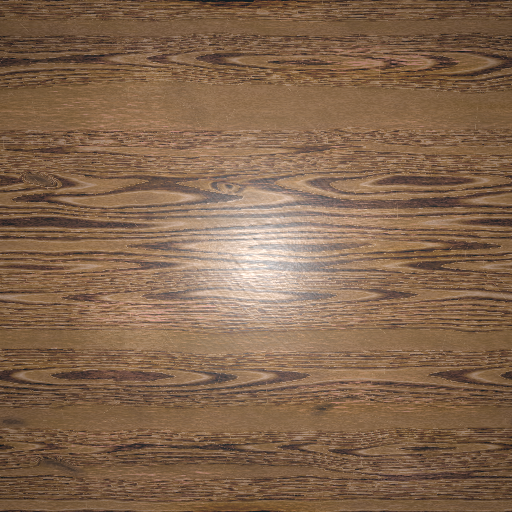} &
\includegraphics[width=0.11\textwidth]{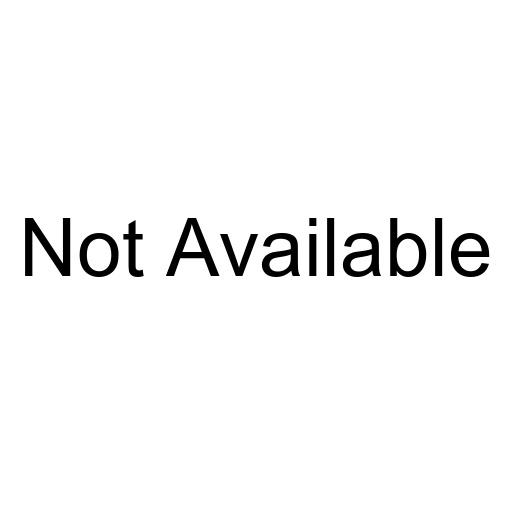} &
\includegraphics[width=0.11\textwidth]{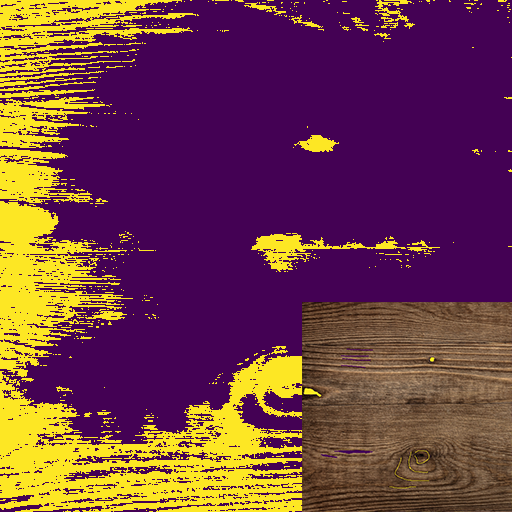} \\
\includegraphics[width=0.11\textwidth]{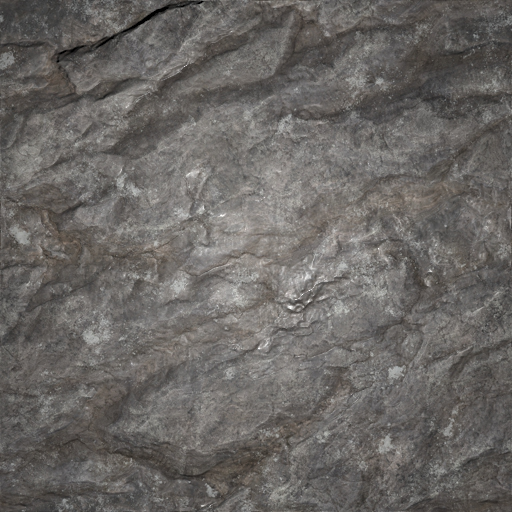} &
\includegraphics[width=0.11\textwidth]{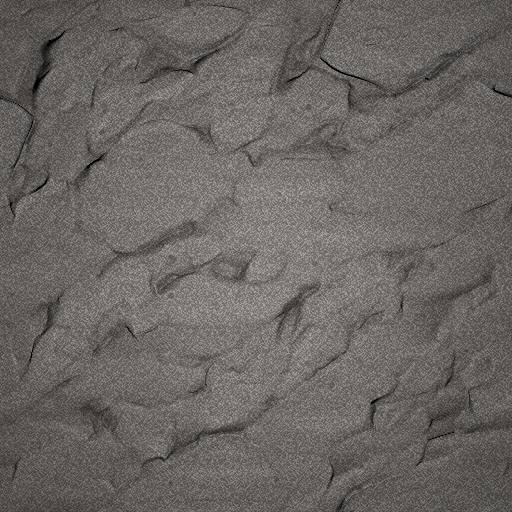} &
\includegraphics[width=0.11\textwidth]{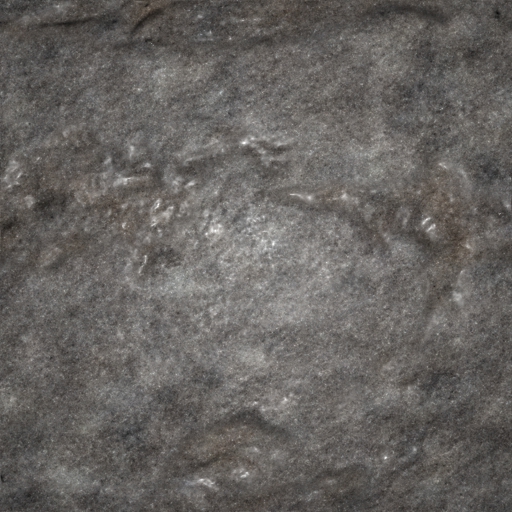} &
\includegraphics[width=0.11\textwidth]{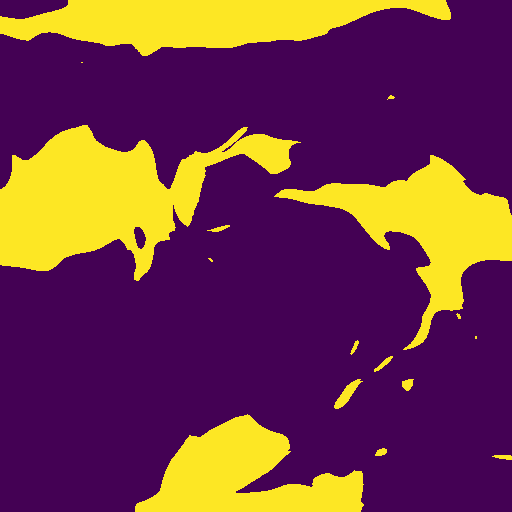} \\
\includegraphics[width=0.11\textwidth]{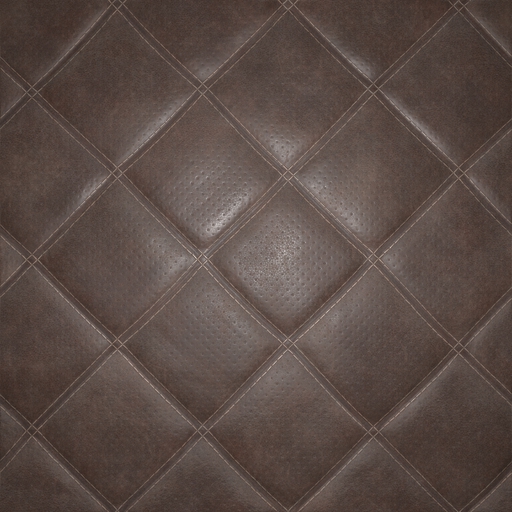} &
\includegraphics[width=0.11\textwidth]{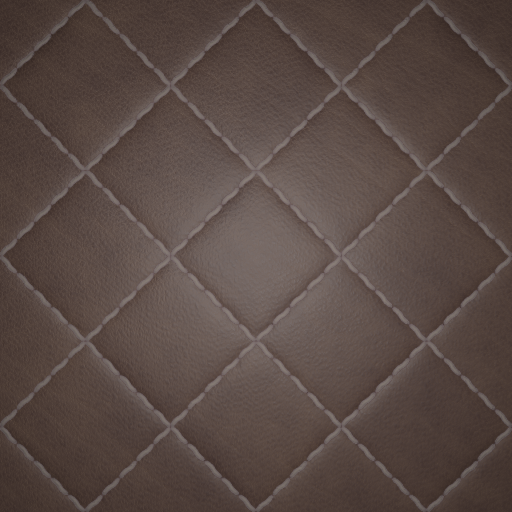} &
\includegraphics[width=0.11\textwidth]{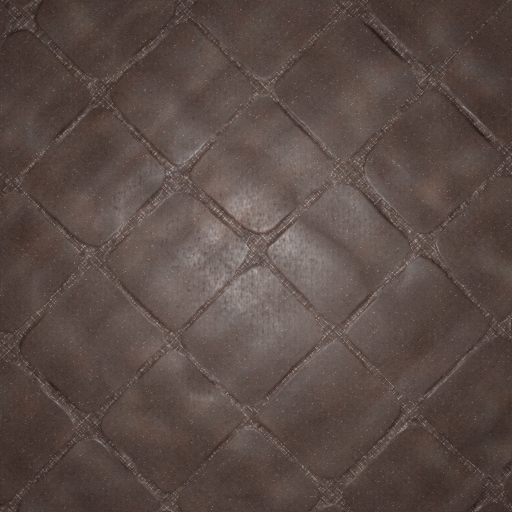} &
\includegraphics[width=0.11\textwidth]{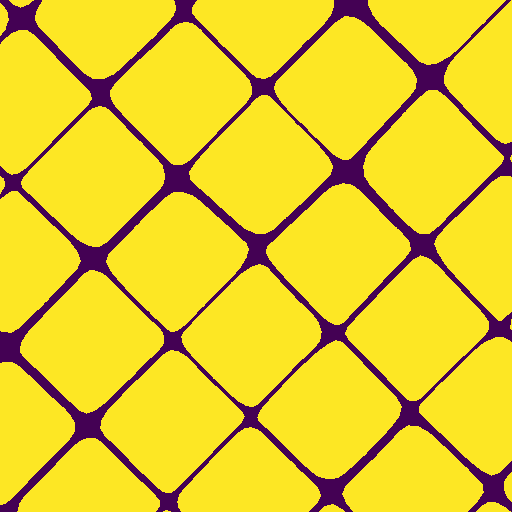} \\
Input & Ours & \multicolumn{2}{c}{Hu et al.'s results and label map}
\end{tabular}
\caption{Comparison to the state-of-the-art inverse procedural modeling method. The semi-automatic pipeline by Hu et al. \shortcite{hu2022inverse} requires segmentation as a starting point. However, not all material maps are easily segmentable e.g, the wood pattern, causing the first step of the algorithm to fail (first row). We show the scribbles used in the matting algorithm (segmentation) overlapped with the albedo map (Inset). Additionally, they model the appearance by approximating the Power Spectrum Density (PSD) of each material using a Gaussian noise, which is not expressive enough to model the appearance of some materials (second row structure and third row appearance).}
\label{fig:comp2hu22}
\end{figure}

Our method can also be considered as an inverse procedural modeling framework if using images as input. We therefore show a comparison with the recent inverse material material modeling framework by Hu et al.~\shortcite{hu2022inverse} in Fig. \ref{fig:comp2hu22}. Compared to their method, our model doesn't require user segmentation and can produce multiple variations rather than a single output graph. More importantly, the gamut of the materials we handle is larger than this previous work, as can be seen in the failure of the previous work for the wood sample in the first row of Fig.~\ref{fig:comp2hu22}. %

\subsection{Multi-modal Conditioning}
We now evaluate our generated procedural material graphs qualitatively. \textbf{For all conditions, we show additional results in Fig. \ref{fig:teaser} and the supplementary materials.} %
\subsubsection{Image Conditioning}
We now show conditionally generated graphs on randomly sampled synthetic images from the test set, where inputs are rendered images with co-located point light. Our generated samples are structurally and semantically close to the input image. We however observe small color mismatch due to prediction errors. Indeed, the color is controlled by very few parameters, which have very strong visual importance, making the inference of the exact color challenging. We use image-space differentiable optimization~\cite{Shi20} to address this issue (Sec. \ref{sec:optim}).

\begin{figure*}
\centering
\includegraphics[width=0.98\textwidth]{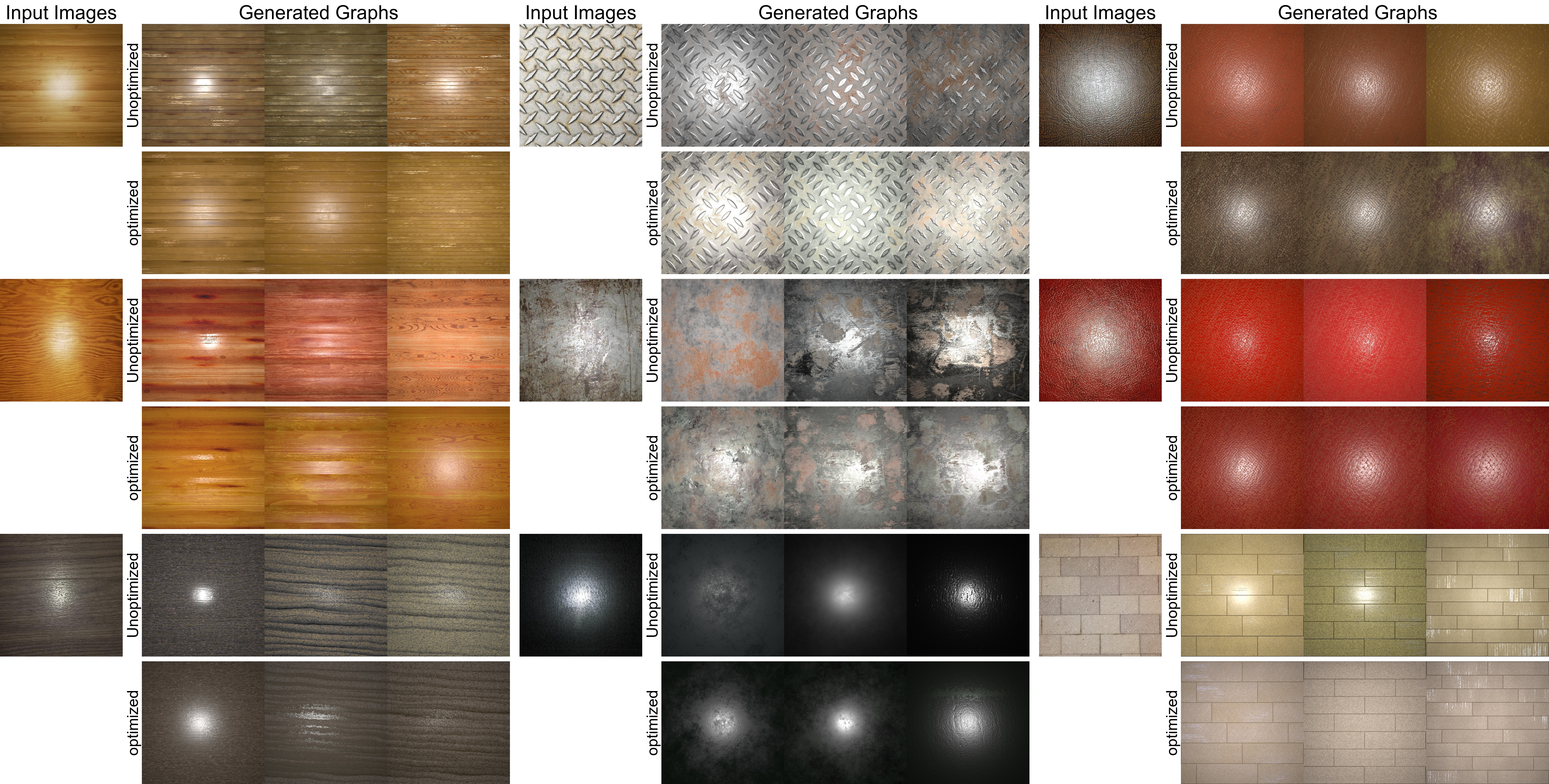}
\caption{Real Image Conditioning. We show real photos as inputs and three of our generated material graphs. For each example, the top row shows non-optimized outputs directly predicted by our models, while the bottom row shows our final output after optimization. See supplemental material for more results.}
\label{fig:image_inputs}
\end{figure*}

Real photographs are more challenging to match as they can only be approximated by procedural models in most cases. %
In Fig. \ref{fig:image_inputs}, we show three generated material graphs before and after optimization when given real photograph inputs. Despite the challenges, our sampled procedural graphs can reproduce the target appearance well, matching its semantics and style. Given our generated graphs, the optimization step is once more capable of adjusting the color and roughness values. %
We also show a comparison to a class-conditioned generation in supplemental material, illustrating the benefits of our CLIP-based conditional generation.

\subsubsection{Text Conditioning}
\begin{figure*}
\centering
\includegraphics[width=0.98\textwidth]{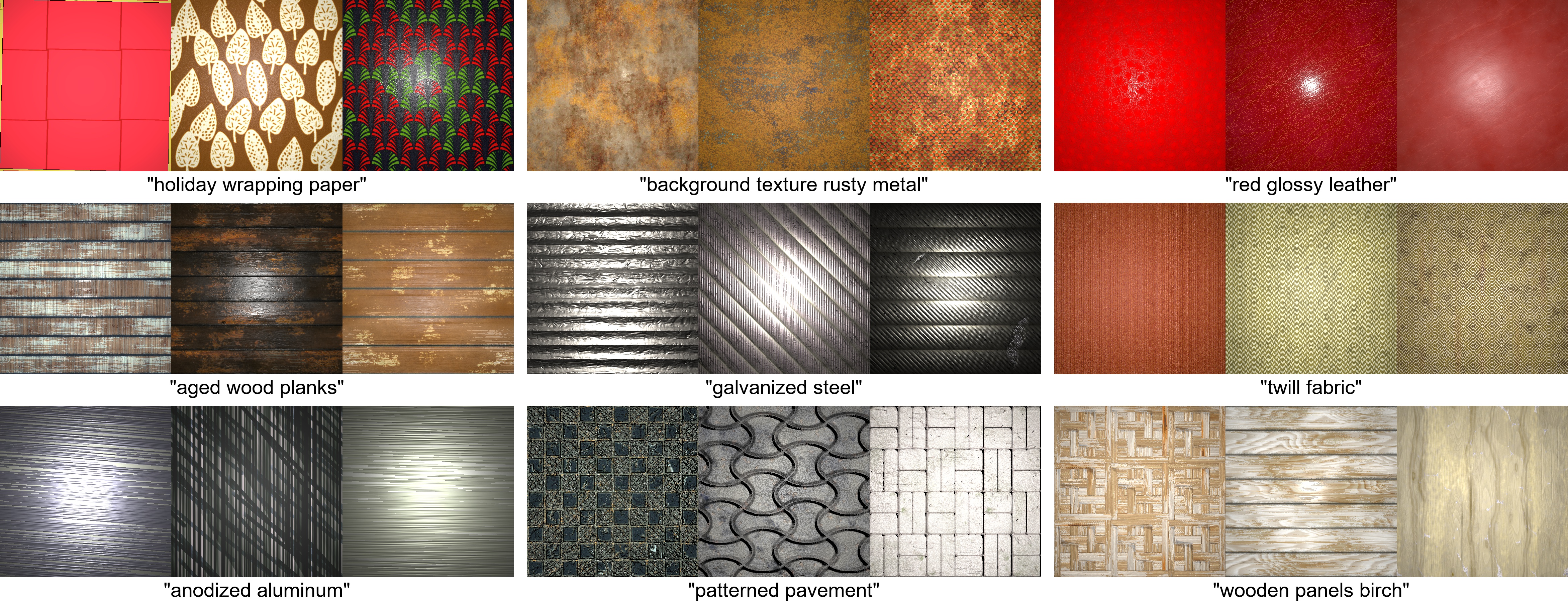}
\caption{Text Conditioning. Our model generates multiple procedural material graphs given various text prompts. See supplemental material for more results.}
\label{fig:text_inputs}
\end{figure*}

As described in Sec. \ref{Sec:cond_gen}, we propose to encode the given text prompt using CLIP and transform the embedding to the CLIP image embedding. We show three generated material graphs for each text prompt in Fig. \ref{fig:text_inputs}. We can see that our model can generate diverse, semantically close material graphs. %

\subsubsection{Unconditional Generation}
\begin{figure}
\centering
\includegraphics[width=0.48\textwidth]{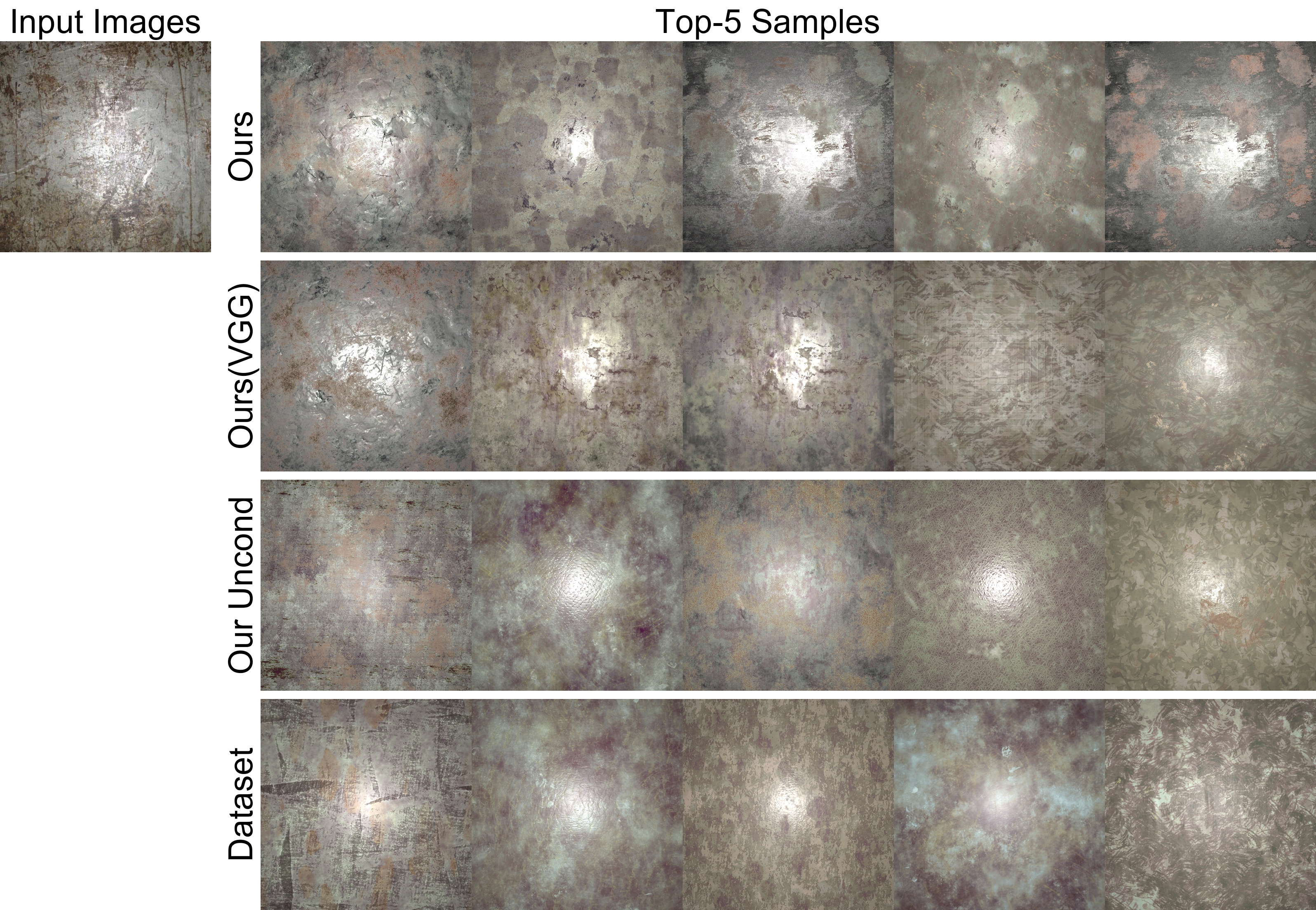}
\caption{We show visual comparisons to our baselines. The top-5 samples are used to calculate the statistics shown in Table \ref{table:stats}. Visually, our model generates material graphs similar to a query in a huge database.}
\label{fig:comp}
\end{figure}

Our augmented dataset and regularized sampling process benefits unconditional material graph generation as well. We present a side-by-side comparison in our supplemental material, showing more diverse samples we generated, compared to MatFormer. We also use this unconditional version to generate the database containing 102,400 as our baseline comparison (\textbf{Ours Uncond} in Table \ref{table:stats} and Fig. \ref{fig:comp})

\subsubsection{Autocompletion: Conditioning on Partial Graphs.}
\label{Sec:autocomplete}
\begin{figure}
\centering
\includegraphics[width=0.47\textwidth]{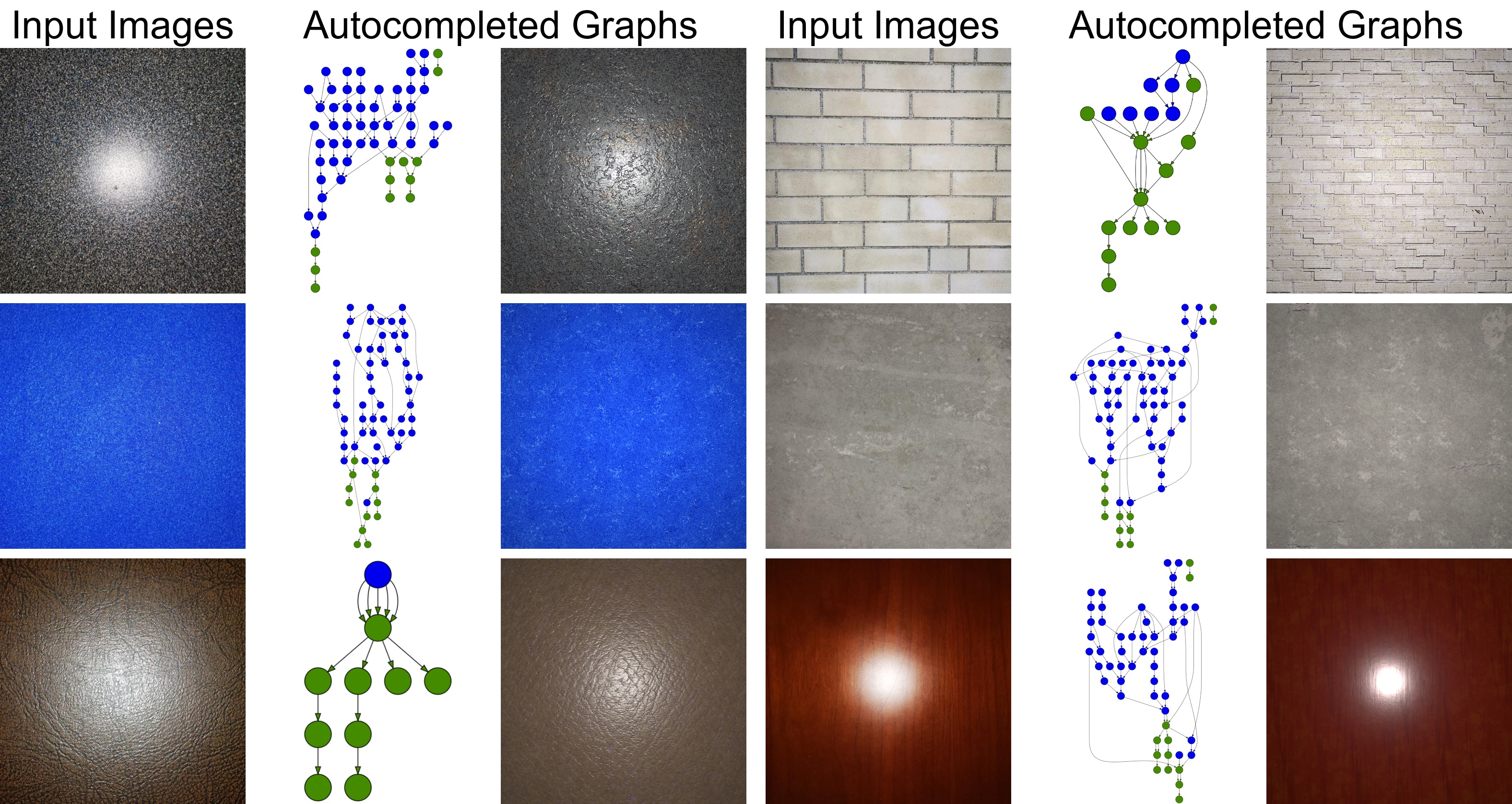}
\caption{As a sequential model, our model can accept partial sequences (partially completed material graphs) and generate the rest of the structures and parameters toward image prompts. As in Fig. \ref{fig:teaser}, existing structures are blue and our predicted are green. See supplemental for more results.}
\label{fig:autocompletion}
\end{figure}

As a sequence-based model, our model can be conditioned on partial graphs inputs and automatically predict the rest of the graph structures and parameters towards a conditional input. While MatFormer is also capable of autocompletion, it cannot be conditioned on a particular desired appearance. For example, in Fig. \ref{fig:autocompletion}, we show examples of automatic visual programming, by generating the end of a partial graph (marked as green nodes and edges) to match target images. This application provides interesting graph modeling exploration possibilities for artists.  

\subsection{Failure Cases and Limitations}
Despite the conditional results we show, limitations remain. The biggest limitation is in the amount of data available. Although we perform data processing and augmentation to build a material dataset $2.32x$ (before filtering) or $1.65x$ (after filtering) larger than previous state-of-the-art \cite{Guererro22}, the scale of our dataset is still far smaller than that of the dataset used for training large language or generative models (\cite{brown2020language, StableDiffusion, Dalle2}. Augmenting the number of base graphs is not trivial, as each is manually crafted by an artist. Our generative model is therefore limited to the subset of appearance and procedural material graphs which artists found useful to design.

\begin{figure}
\centering
\includegraphics[width=0.47\textwidth]{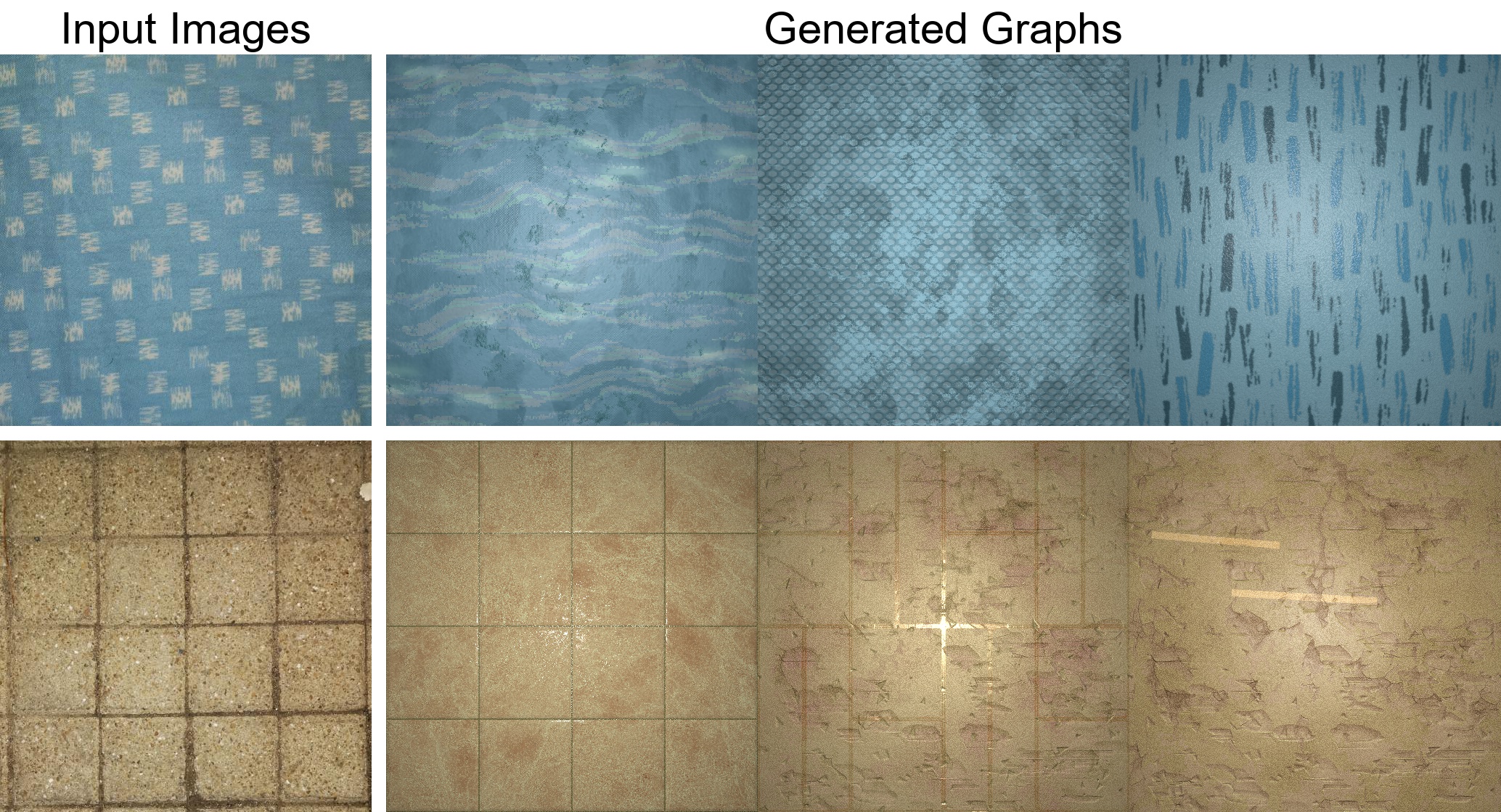}
\caption{
Our method is unable to reproduce the detailed appearance of some image inputs due to limited training data and prediction errors.}
\label{fig:fail_im}
\end{figure}

\begin{figure} [hp]
\centering
\includegraphics[width=0.47\textwidth]{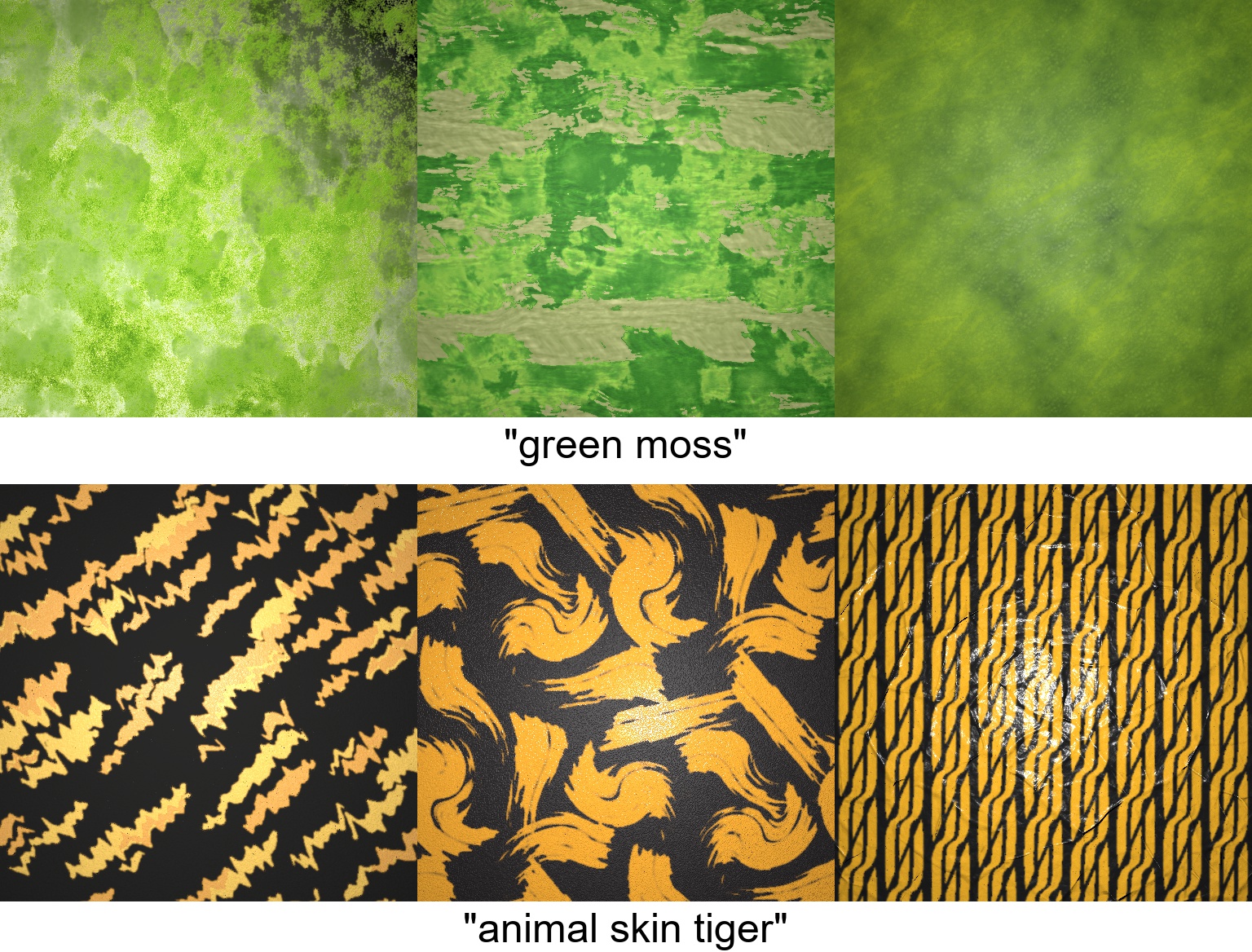}
\caption{Material graphs sampled by text prompts do not always generate realistic materials which match the expected appearance given the text prompt.}
\label{fig:fail_text}
\end{figure}

In Figs. \ref{fig:fail_im} and \ref{fig:fail_text}, we show failures cases of our generative model where the generated material graphs do not match well the input prompts for images/texts. While the overall structure is correct, the details do not match. Further, our model currently only supports standard PBR workflow material maps (Base Color, Normal, Roughness, Metalness) as they are the most represented in the dataset. Finally, the quantization step for token prediction is a limitation. While we choose a finer quantization than previous work, quantization errors still happen, and predicting a continuous field rather than discrete bins would be an interesting future work.
\section{Conclusion and Future Work}
We present the first conditional generative model for procedural material node graph generation. We show that our model generates high-quality node graphs given either images or text prompts. The proposed generative model is a new tool for users to explore the design space of materials and has interesting applications such as automatic visual programming (Sec. \ref{Sec:autocomplete}).

As the dataset is a crucial component, an interesting future work would be to expand it. But more interestingly, the quality of the graphs could be improved. We take a first step in this direction with our cleanup step, but beyond cleanup, the existing material graphs designed by artists do not follow standard programming design principles. In particular, they are not necessarily intended to be easily modified or reused. 
An interesting future step would be to find reuseable sub-graphs i.e., sub-functions and expand the dataset, using a set of hierarchically organized modules or libraries. This would both reduce the sequence length and create more structure.

Finally, there is a gap between predicted token space error and image/text space matching error. We currently minimize this gap using an image/text-space optimization as a post-processing step, but further improvement would require finding a way to evaluate and minimize the image/text space error during training and inference.

\begin{acks}
This work was supported in part by NSF Grant No. IIS-2007283.
\end{acks}

\bibliographystyle{ACM-Reference-Format}
\bibliography{bibliography}
\clearpage
\begin{figure*}
\centering
\includegraphics[width=0.99\textwidth]{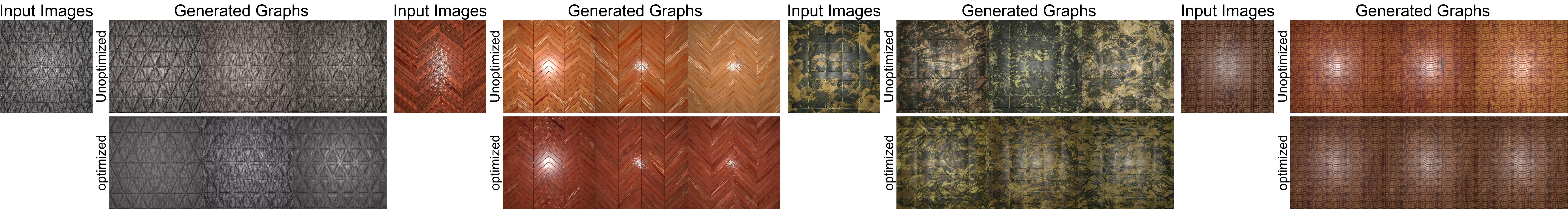}
\caption{Synthetic Image Conditioning. We show three generated procedural material for each input. For each example, the top row shows non-optimized outputs directly predicted by our models, while the bottom row shows our final output after optimization, which easily rectifies the colors.}
\label{fig:synthetic}
\end{figure*}

\clearpage
\clearpage

\end{document}